\documentclass[sigconf]{acmart}
\setcopyright{none}
\renewcommand\footnotetextcopyrightpermission[1]{} % removes footnote with conference information in first column

\usepackage{multirow}
\usepackage{caption}

\usepackage{amssymb}
\usepackage{amsmath}
\usepackage{subcaption}
\usepackage{graphicx}
\usepackage{float} 
\usepackage{enumitem}
\usepackage{algorithm} 
\usepackage{algorithmic}
\usepackage{booktabs} 
\usepackage{threeparttable}
\usepackage{makecell}
\usepackage{balance}
\usepackage{xspace}

\newcommand{\model}{UniFormer\xspace}
\AtBeginDocument{%
  \providecommand\BibTeX{{%
    \normalfont B\kern-0.5em{\scshape i\kern-0.25em b}\kern-0.8em\TeX}}}

\copyrightyear{2027} 
\acmYear{2027} 
\setcopyright{cc}
\setcctype{by}
\acmDOI{10.1145/3711896.3737242}
\acmISBN{979-8-4007-1454-2/2025/08}

\begin{document}

\title{UniFormer: Efficient and Unified Model-Centric Scaling for Industrial Recommendation}

% \titlenote{Co-first authors with equal contributions.}

\newcommand{\chenxu}[1]{{\bf \color{red} [[Chenxu says ``#1'']]}}

% Put this before \maketitle
% \settopmatter{authorsperrow=4}

\author{
Bo Chen$^{*}$,
Jinlong Jiao$^{*}$,
Tijian Hu$^{*}$,
Ruihao Zhang,
Yanzhi Liu,
Chenghou Jin,
Qinglin Jia,
Baixuan He,
Wenhao Li,
Hechang Pan,
Yiwu Liu,
Jian Liang,
Chaoyi Ma$^{\dagger}$,
Ruiming Tang$^{\dagger}$,
Han Li,
Kun Gai
}

\affiliation{%
  \institution{Kuaishou Technology}
  \city{Beijing}
  \country{China}
}

\email{{renze03, jiaojinlong, hutijian, machaoyi03, tangruiming}@kuaishou.com}

\thanks{$^{*}$ Equal contributions. \\$^{\dagger}$ Corresponding authors.}

\renewcommand{\shortauthors}{Bo Chen et al.}

% \renewcommand{\shortauthors}{xxx, et al.}

\begin{abstract}
Recently, substantial progress has been made in industrial recommendation through component-centric model scaling, where individual components such as behavior modeling, feature interaction, or task modeling are independently scaled to improve model capacity.
Although recent methods such as HyFormer and OneTrans further explore cross-module co-scaling by jointly modeling behavior and interaction, their designs are still confined to the feature space and lack a unified model-centric scaling framework over the overall modeling space.
In this paper, we propose \textbf{\model}, an efficient and unified model-centric scaling framework for industrial recommender systems. To improve efficiency, \model decomposes the overall modeling space into feature and task spaces, which are modeled by stacked Feature-space Interaction Modules and Task-space Interaction Modules, respectively. 
Moreover, \model introduces semantic-based tokenization scheme to enable user-item decoupling, thereby achieving request-level inference acceleration.
To prevent attention collapse, \model employs multi-sequence cross-attention to separately capture heterogeneous behavior patterns, followed by the self-attention to enhance interaction modeling. 
Besides, dedicated multi-view FFNs are introduced to support flexible and scalable parameter scaling across different modeling components. 
% Extensive online A/B testing in two large-scale short-video recommendation scenarios shows that \model consistently improves user engagement and interaction metrics, achieving gains of +0.102\%/+0.260\% in App Stay Time and +0.729\%/+1.113\% in Watch Time, respectively.
% 匿名
Extensive online A/B testing in two production scenarios, Kuaishou and Kuaishou Lite, shows that \model consistently improves user engagement and interaction metrics, achieving gains of +0.102\%/+0.260\% in App Stay Time and +0.729\%/+1.113\% in Watch Time, respectively.

\end{abstract}








\maketitle

\section{INTRODUCTION}\label{sec:intro}
% 介绍工业推荐系统
% 工业推荐系统的特点
% NLP社区通过扩大模型/数据/计算效果取得了效果的显著提升。REc社区开始围绕推荐系统的特点尝试通过scaling 提升推荐效果。

% 二、点出之前推荐系统scaling 考虑的都是单方面的、没有考虑co-scaling
% 近期的推荐系统在scaling方向的研究工作可以分为三类、

% 3、2、1.
% 1、采用moe的方案进行多任务的建模。（SMES、MSN）（介绍一下做法）
% 2、采用类似transformer的结构进行高阶的特征交叉。（mixer，unimier）(介绍一下做法)
% 3、主要是利用transformer建模序列特征（longer stca）。（介绍一下做法）

% 三、problem / challenge
% CH1：这些方向各自独立的发展，使得模型变得非常的碎片化，也不适合工程优化

% NLP社区所有的token使用一样的embedding space，所有的交互都使用transformer完成。
% 有一些工作探索co-scaling feature space【onetrans、hyformer】，但是忽略了任务空间。我们夹的参数在2个空间都能收益。

% CH2：3里面的行为异构的 ，对于1的contribution ，分段式的搞法，给1的建模。且中间信息不产生交互也会损失信息。

% CH3：为什么之前会独立发展的（不同的业务有自己的业务特点），推荐场景很丰富的一个特点，我们能够灵活的设计针对业务特点的 高度自定义的参数分配方案。

\begin{figure}[t]
  \centering
  \includegraphics[width=0.95\linewidth]{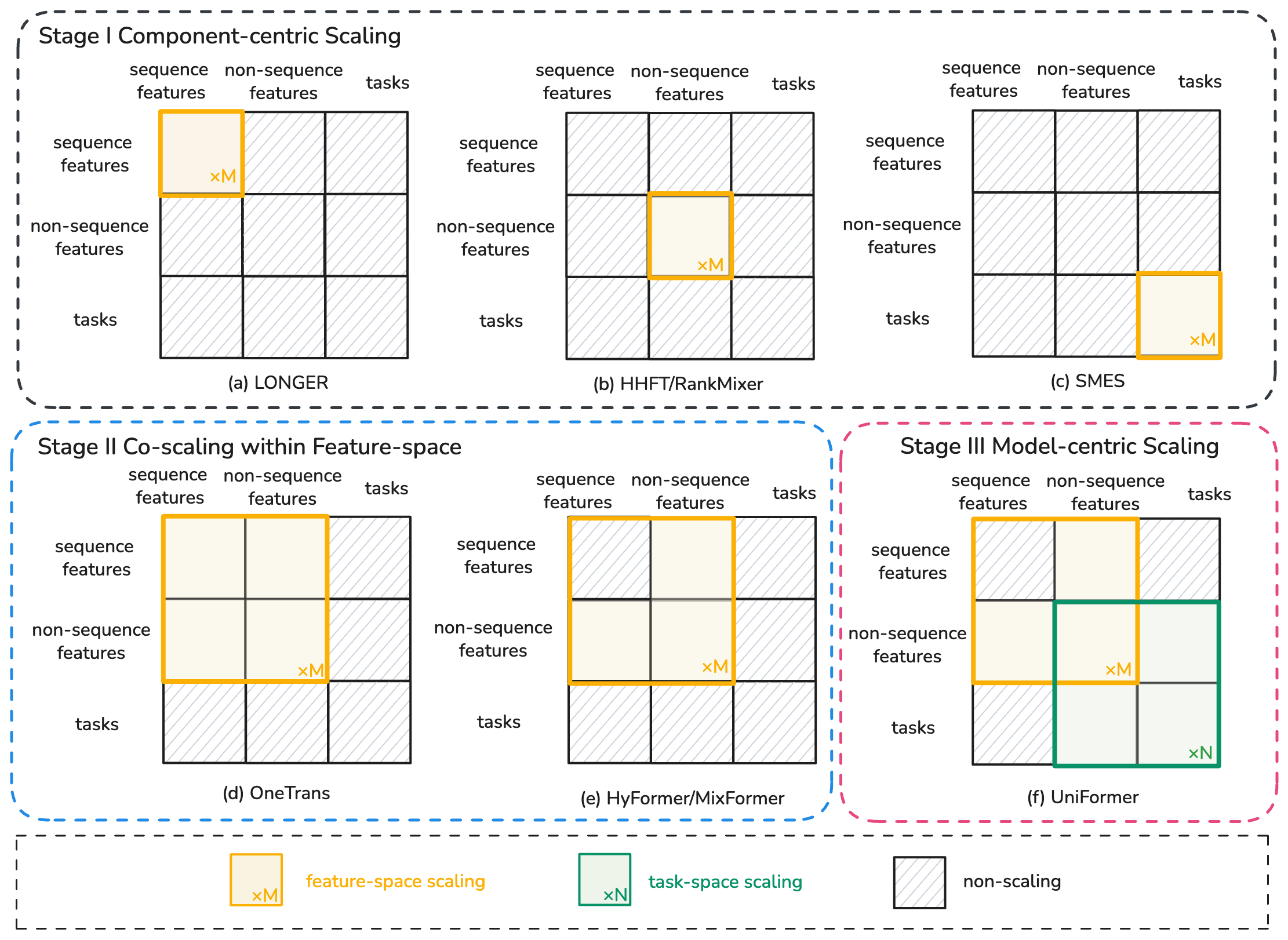}
  \caption{Comparison of representative scaling methods.
  (a)-(c) present the component-centric scaling; (d)-(e) jointly model sequential behavior and feature interaction within the feature space;
%Existing methods mainly scale within the feature space: RankMixer focuses on feature interaction, while OneTrans and HyFormer/MixFormer jointly model sequential behavior and feature interaction. 
(f) \model proposes a unified co-scaling framework, enabling co-scaling across features and tasks.}
  \label{fig:model_comp}
  \vspace{-10pt}
\end{figure}

Industrial recommender systems leverage life-long user behavior histories and heterogeneous features (\textit{e.g.}, user profiles and item metadata) to estimate the utility of various user-item interactions.
For instance, short video sharing platforms (\textit{e.g.}, Kuaishou and TikTok) predict a wide range of user feedback, such as \textit{Click, Like, Share, Comment}, and \textit{Completion}, to determine final video exposure~\cite{rankmixer,smes}.
To manage this complexity, industrial recommendation models typically comprise several modular components built upon the embedding layer: a \textbf{behavior modeling module} for sequential pattern extraction~\cite{din,dien}, a \textbf{feature interaction module} for cross-feature dependencies modeling~\cite{deepfm,dcn,edcn}, and a \textbf{task modeling module} for multi-objective optimization~\cite{mmoe,ple,survey_lbm}.
To this end, a variety of fragmented, hand-crafted operators have been developed~\cite{eenet}. Despite their effectiveness in improving predictive performance, these specialized designs complicate model optimization and limit the utilization of computational resources~\cite{wukong,hstu}.

%The recent success of Large Language Models (LLMs) has established scaling laws as the dominant paradigm for achieving superior performance, thereby driving industrial recommender systems to expand model capacity aggressively.
% Industrial recommendation models leverage long-range user behavior histories and heterogeneous features (user profiles, contextual signals, item metadata) to estimate the utility of various user-item interactions.
% For instance, short video sharing platforms predict a wide range of user feedback—such as liking, sharing, commenting, watch time, and long-viewing, to determine final video exposure.
% To manage this complexity, industrial ranking models typically comprise several modular components: an embedding layer for tokenization, a behavior modeling module for sequential patterns, a feature interaction module for cross-feature reasoning, and a task modeling module for multi-objective optimization.

The remarkable success of large language models (LLMs) has demonstrated the effectiveness of model scaling in achieving superior performance~\cite{deepseek}, motivating industrial recommender systems to scale up their model capacity~\cite{gr_survey,ctrl}.
For \textbf{feature space} model scaling, transformer-based architectures such as HSTU~\cite{hstu}, LONGER~\cite{longer}, and STCA~\cite{stca} improve long-sequence behavior modeling, while HHFT~\cite{hhft} and RankMixer~\cite{rankmixer} scale feature interaction capacity through self-attention or token-mixing operators. Meanwhile, MoE-based methods such as SMES~\cite{smes} enhance multi-task modeling via scalable experts in \textbf{task space}.
Despite their effectiveness, these approaches primarily follow a component-centric scaling paradigm as illustrated in Figure~\ref{fig:model_comp} (a)-(c), which limits the efficient utilization of computational resources and prevents the model from fully realizing the benefits of holistic scaling.

% To further enhance recommendation performance, industrial practices have evolved towards scaling up these modular components.
% However, this scaling effort has been largely disjointed. Rather than evolving as a unified architecture, these components have pursued distinct optimization paths.
% Correspondingly, these scaling efforts manifest in three dominant paradigms: Transformer-based long sequence modeling (LONGER, STCA),
% TokenMixer-based heterogeneous feature interaction (Rankmixer,Unimixer), and Mixture-of-Experts (MoE)-based multi-objective modeling (MSN,SMES).
% In practice, these modules are stacked hierarchically. 
% Nevertheless, current scaling strategies remain largely component-centric. While increasing model capacity or sequence length yields improvements within isolated modules, it fails to yield holistic representations across the architecture. As a result, the overall system exhibits low scaling efficiency, as gains in one area do not translate to proportional improvements in recommendation performance.

Therefore, to fully realize the potential of model scaling for recommendation, the scaling paradigm should evolve \textbf{from Component-centric Scaling toward Model-centric Scaling}. Recent studies, such as OneTrans~\cite{onetrans} and MixFormer~\cite{mixformer}, have made preliminary attempts to explore cross-module co-scaling. However, these approaches remain largely confined to the feature space (illustrated in Figure~\ref{fig:model_comp} (d)-(e)), jointly scaling behaviors modeling and feature interaction while leaving task modeling underexplored.

% The root cause of this inefficiency lies in the historical fragmentation caused by business heterogeneity. Industrial recommender systems serve different business domains, each requiring specialized modeling techniques and optimization goals. To maximize local performance, the industry has long relied on hand-crafted parameter allocation schemes tailored for each scenario. While this specialized approach offers the flexibility needed to address diverse feature distributions and user behaviors, it inevitably leads to significant model fragmentation, preventing the system from achieving proportional improvements through parameter expansion.

To enable efficient and unified model-centric scaling, several key challenges need to be addressed.
First, due to the large number of features and tasks in industrial recommendation systems, as well as their high-concurrency and low-latency serving requirements, the unified co-scaling model must maintain \textbf{high training and inference efficiency} to mitigate the resource overhead introduced by scaling up model capacity~\cite{tokenformer}.
Second, the unified co-scaling model should effectively capture heterogeneous user behaviors and short/long-term interests, ensuring comprehensive personalization and \textbf{preventing attention collapse}~\cite{gems}.
Third, the unified co-scaling model should support \textbf{flexible and scalable parameter allocation}, enabling balanced capacity scaling across different components and avoiding excessive parameter concentration in any single module~\cite{tokenmixer}.

To address the above challenges, we propose \textbf{\model}, an efficient and unified model-centric scaling framework designed for industrial recommender systems.
Specifically, to improve efficiency, \model decomposes the unified architecture into feature-space interaction (FIM) and task-space interaction (TIM), as illustrated in Figure~\ref{fig:model_comp} (f). The former scales the modeling of sequential and non-sequential features, while the latter captures interactions between high-order features and task-specific signals. 
By omitting the costly intra-sequence interactions (similar to the lazy design principle of OneRec-v2~\cite{onerec_v2}) and interactions between sequences and tasks, \model achieves substantial efficiency improvements.
Besides, a semantic-based tokenization scheme is proposed, enabling user-item decoupling during inference and achieving request-level inference acceleration.
To prevent attention collapse, FIM employs a multi-sequence cross-attention mechanism to model heterogeneous behavior sequences separately, followed by self-attention to enhance interaction modeling within the feature space.
Moreover, to enable flexible and scalable parameter allocation, \model introduces multi-view Feed-Forward Networks (FFNs) for sequential features, non-sequential features, and task signals, respectively, thereby supporting balanced capacity scaling.

In summary, our main contributions are summarized as follows:
\begin{itemize}[leftmargin=*]
\item We advocate a paradigm shift from component-centric scaling to model-centric scaling for recommendation models and systematically identify the key challenges in building an efficient and unified scaling framework.
\item We propose \textbf{\model}, an efficient and unified model-centric scaling framework that integrates Feature-space Interaction (FIM) and Task-space Interaction (TIM) modules, leveraging standardized attention operators and multi-view FFNs to enhance modeling capacity while enabling flexible and scalable parameter allocation.
% \item We conduct extensive offline experiments and analyses in a large-scale short-video recommendation scenario, demonstrating the effectiveness of \model. Furthermore, online A/B testing in the production system shows consistent improvements in user engagement and interaction metrics.
% 匿名
\item We conduct extensive offline experiments and analyses in Kuaishou's industrial recommendation scenarios, demonstrating the effectiveness of \model. Furthermore, online A/B testing in two production scenarios, Kuaishou and Kuaishou Lite, shows consistent improvements in user engagement and interaction metrics.

\end{itemize}

\section{RELATED WORK}\label{sec:related_word}
\subsection{Pre-scaling Recommendation Models}
Deep learning-based recommendation models typically consist of two major components: sparse embedding tables and dense neural networks (DNNs). During the pre-scaling era, the majority of model parameters were concentrated in sparse embedding tables, while the DNNs remained relatively lightweight, resulting in a characteristic sparse-embedding-dominant design~\cite{deepfm,dcn,inttower,stem}. 
The dense modeling part was further constructed in a modular manner with independently designed components for behavior modeling, feature interaction, and task modeling. %Consequently, improvements were primarily driven by component-level architectural innovations rather than by scaling the dense modeling capacity.

For behavior modeling, which focuses on capturing user interests from historical interaction sequences, DIN~\cite{din} and DIEN~\cite{dien} leverage attention mechanisms to model the relevance between user behaviors and target items. To model lifelong user interests, search-based methods such as SIM~\cite{sim} and TWIN~\cite{twin} retrieve the most relevant behavior subsequences from long behavior sequences. In contrast, VISTA~\cite{vista} and C-Former~\cite{cformer} adopt compression-based paradigms to extract compact interest representations, enabling efficient modeling of lifelong user preferences.
For feature interaction modeling, methods such as DeepFM~\cite{deepfm}, DCN~\cite{dcn}, EENet~\cite{eenet}, and their various extensions aim to capture relationships among heterogeneous features through different interaction operators, including inner product, outer product, and attention mechanisms. These models improve recommendation performance by explicitly or implicitly modeling high-order feature interactions within representation spaces.
For task modeling, multi-task recommendation models such as MMoE~\cite{mmoe}, PLE~\cite{ple}, and HoME~\cite{home} improve overall multi-task learning capability by designing various task-shared and task-specific network architectures to balance knowledge sharing and task specialization across multiple objectives.
Despite their effectiveness, these models mainly optimize individual modules independently and rely on relatively small-scale DNNs, without exploring parameter scaling across different modules~\cite{wukong,hstu}.

%Traditional deep learning-based recommendation models mainly focus on improving specific components of the recommendation pipeline, while the overall model scale remains relatively limited. Existing approaches can generally be categorized into three directions: behavior modeling, feature interaction modeling, and task modeling. Behavior modeling methods aim to capture user interests from historical behaviors, including attention-based methods such as DIN, retrieval-based methods such as SIM and TWIN, and long-sequence modeling methods such as LONGER. Feature interaction modeling methods focus on learning high-order interactions among heterogeneous features, including factorization-based approaches such as DeepFM, explicit cross-feature modeling methods such as DCN, and unified feature interaction frameworks such as WUKONG. Task modeling methods further extend recommendation models to multi-task settings, where representative approaches include MMOE, PLE, and HOME, which improve performance through task-specific expert routing and shared representation learning. Despite their effectiveness, these methods are typically designed for relatively small-scale DNN architectures and mainly optimize individual modeling components independently, without exploring unified scaling across behavior modeling, feature interaction, and task modeling.

\subsection{Scaling Recommendation Models}
Recently, parameter scaling in large language models (LLMs) has demonstrated remarkable emergent capabilities in world knowledge and reasoning, inspiring the evolution of recommendation models toward scaling paradigms. 
Recent recommendation architectures mainly focus on component-centric scaling.
Transformer-based architectures such as HSTU~\cite{hstu}, PinFM~\cite{pinfm}, LONGER~\cite{longer}, and STCA~\cite{stca} enhance long-sequence behavior modeling, while HHFT~\cite{hhft} and RankMixer~\cite{rankmixer} model feature interaction through self-attention or token-mixing. MoE-based methods such as SMES~\cite{smes} introduce scalable experts for multi-task learning.

To enable unified modeling of sequential and non-sequential features, OneTrans~\cite{onetrans} and TokenFormer~\cite{tokenformer} adopt self-attention architectures over the entire feature space. They further incorporate pyramid-style progressive shrinking and sliding-window attention to mitigate the computational overhead.
Furthermore, works such as HyFormer~\cite{hyformer}, MixFormer~\cite{mixformer}, HeMix~\cite{hemix}, and EST~\cite{est} design more efficient interaction paradigms, where interactions between sequential and non-sequential features are modeled through lightweight cross-attention, while feature interactions are handled via token mixing or self-attention, thereby avoiding the substantial computational cost introduced by full self-attention.
However, existing unified co-scaling recommendation models primarily focus on feature-space interaction modeling,  with limited exploration of unified modeling across feature and task spaces.
%Moreover, heterogeneous sequence modeling and efficient parameter scaling remain insufficiently investigated.

% \newpage

\section{Preliminaries}\label{sec:preliminaries}
Deep learning-based industrial recommendation models typically consist of several modular components, including an embedding (\textit{i.e.}, tokenization) layer, a behavior modeling module, a feature interaction module, and a task modeling module~\cite{autodis,smes}. Given input features $x$, the embedding layer produces dense representations $\textbf{e}=\phi(x)$~\cite{autodis}. The overall process can be formulated as follows:
\begin{equation}
\mathbf{y} = \mathcal{F}_{\text{task}}\Big( 
    \mathcal{F}_{\text{interaction}}\big( 
        \mathcal{F}_{\text{behavior}}(\mathbf{e}_{\text{seq}}),\; 
        \mathbf{e}_{\text{non\text{-}seq}} 
    \big),\mathbf{e}_{\text{task}}
\Big),
\end{equation}
where $\mathcal{F}_{\text{behavior}}(\cdot), \mathcal{F}_{\text{interaction}}(\cdot),$ and $\mathcal{F}_{\text{task}}(\cdot)$ denote behavior modeling (\textit{e.g.}, DIN and SIM)~\cite{din,sim}, feature interaction (\textit{e.g.}, DeepFM and DCN)~\cite{deepfm,dcn,edcn}, and task modeling functions (\textit{e.g.}, MMoE and PLE)~\cite{mmoe,ple,multi_task}, respectively. Despite their effectiveness, these modules are typically designed independently, resulting in a fragmented architecture that limits unified modeling~\cite{wukong,hstu}.

Recently, with the rapid advancement of large language models (LLMs) and the significant increase in computational power, model scaling has emerged as a key factor driving the emergence of powerful capabilities across various domains. Inspired by this trend, recommendation models have also begun to explore scaling strategies. %Existing approaches mainly focus on scaling within the feature modeling space~\cite{rankmixer,mixformer,onetrans}, as illustrated in Figure~\ref{fig:model_comp}. %For example, RankMixer~\cite{rankmixer} and HHFT~\cite{hhft} emphasize scaling feature interaction by leveraging token mixing or self-attention mechanisms. Furthermore, works such as OneTrans~\cite{onetrans} and HyFormer~\cite{hyformer} extend this paradigm by jointly scaling behavior modeling and feature interaction. %Specifically, OneTrans adopts full self-attention over all features, while HyFormer decomposes this process into cross-attention for modeling interactions between sequential and non-sequential features and self-attention for (aggregated) non-sequential feature interaction, thereby improving efficiency.
However, these methods primarily focus on scaling within the feature modeling space~\cite{rankmixer,mixformer,onetrans}, overlooking the role of task modeling in the scaling paradigm. %In practice, recommendation systems often involve multiple tasks (e.g., click-through rate, conversion rate), where task-specific objectives introduce additional modeling complexity beyond feature interactions. 
To address this limitation, we propose a unified scaling framework that extends beyond feature space and incorporates task modeling as a first-class component:
\begin{equation}
\mathbf{y} = \mathcal{F}_{\text{co-scaling}}\Big( 
    \mathbf{e}_{\text{seq}}, 
        \mathbf{e}_{\text{non-seq}},\mathbf{e}_{\text{task}} 
\Big).
\end{equation}
% By jointly scaling sequential behavior modeling, feature interaction, and task modeling in a unified manner, our approach enables more expressive and efficient representation learning.

\section{Methodology}\label{sec:method}
%In this section, we first introduce the overall architecture of the proposed model \model. We then present the two key components, tokenization and unified interaction, in detail. Finally, we describe the optimization and deployment of the \model.
\subsection{Overall} \label{sec:method-overall}
\begin{figure*}[t]
  \centering
  \includegraphics[width=0.7\linewidth]{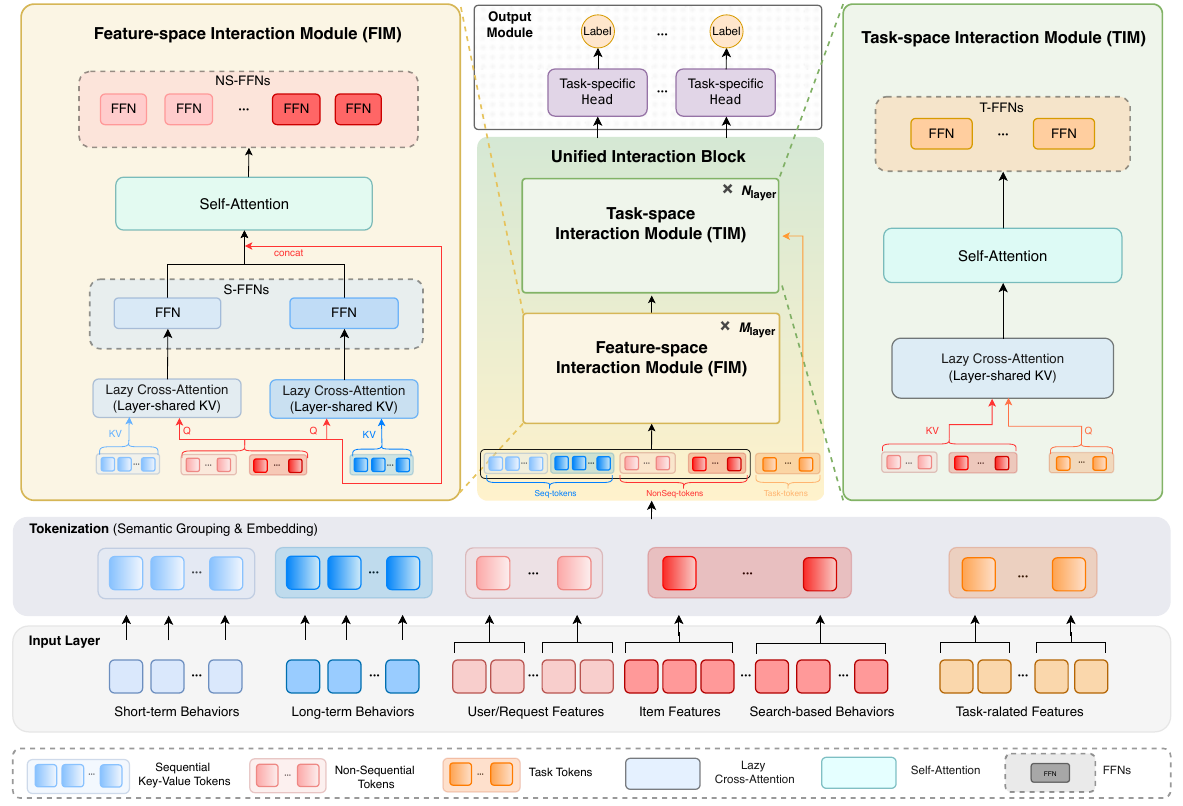}
  \caption{\textbf{Overall architecture of \model}, which consists of two core components: tokenization and the unified interaction block. The unified interaction block is composed of stacked Feature-space Interaction Modules (FIMs) and Task-space Interaction Modules (TIMs). FIM takes tokenized sequential and non-sequential features as input and captures feature-space interactions through attention and FFNs layers. TIM further takes the high-order representations produced by FIM together with tokenized task features, and adopts the same standardized operators to capture task-feature and inter-task interactions.}
  \label{fig:model}
  \vspace{-10pt}
\end{figure*}

Figure~\ref{fig:model} illustrates the proposed unified co-scaling framework \model, which consists of two core components: a tokenization block and a unified interaction block. The tokenization block transforms sequential features, non-sequential features, and task-related features from the input layer into compact token representations through feature grouping and projection. To avoid the prohibitive computational overhead of modeling interactions across the entire space with full attention (exemplified by OneTrans) and achieve efficient model-centric co-scaling, \model decouples the overall modeling space into feature and task spaces, which are modeled by stacked Feature-space Interaction Modules (FIMs) and Task-space Interaction Modules (TIMs), respectively. Each module follows a standard Transformer-style architecture composed of attention layers and FFNs, enabling scalable parameter allocation while maintaining computational efficiency.
% Specifically, FIM focuses on scaling within the feature space. It first employs multi-sequence cross-attention and sequence-specific FFNs (S-FFNs) to capture heterogeneous user interests from multiple behavioral perspectives, and then applies self-attention and non-sequential FFNs (NS-FFNs) to model high-order feature interactions. TIM extends a similar design to the task space, efficiently capturing task-feature and inter-task interactions for scalable multi-task modeling.

\subsection{Tokenization} \label{sec:method-tokenization}

\subsubsection{Sequential Features}\label{sec:method-tokenization-seq}
User behavioral sequence features can be categorized into two types based on their dependency on the target item. The first type is item-independent behavior, including short-term interaction sequences and long-term compressed interest representations (\textit{e.g.}, C-Former~\cite{cformer}). The second type is item-dependent behavior, such as candidate-aware search-based sequences (\textit{e.g.}, SIM~\cite{sim}). To effectively model these heterogeneous patterns, we design distinct tokenization strategies.

For \textbf{item-independent behaviors}, we preserve the original sequential structure and tokenize each element in the sequence individually. For short-term interaction sequences, the $i$-th element is represented by concatenating the item's side information, such as video ID embeddings $\mathbf{e}^{\mathrm{pid}}_{i}$, content type $\mathbf{e}^{\mathrm{tag}}_{i}$, creator information $\mathbf{e}^{\mathrm{paid}}_{i}$,  watch duration $\mathbf{e}^{\mathrm{duration}}_{i}$ and so on, forming a token representation $\mathbf{e}^{\mathrm{short}}_{i} = [\mathbf{e}^{\mathrm{pid}}_{i}||\mathbf{e}^{\mathrm{tag}}_{i}||\mathbf{e}^{\mathrm{paid}}_{i}||\mathbf{e}^{\mathrm{duration}}_{i}||\dots]$. For long-term compressed interest representations, we directly retain the pretrained embedding without further modification, i.e,  $\mathbf{e}^{\mathrm{long}}_{i} = \mathbf{e}_{i}^{\mathrm{pretrained}}$.

Subsequently, following OneRec-v2~\cite{onerec_v2}, each representation $\mathbf{e}_{i}$ of different sequences is projected into a unified hidden space via SwiGLU layer~\cite{tokenmixer}, yielding $\mathbf{z}_{i} \in \mathbb{R}^{d_{\mathrm{seq}}}$ with dimension:
\begin{equation}
d_{\mathrm{seq}} = S_{\mathrm{kv}} \cdot L_{\mathrm{kv}} \cdot d_{\mathrm{model}},
\end{equation}
where $S_{\mathrm{kv}}$ is the key-value split coefficient, and $L_{\mathrm{kv}}$ is the number of key-value layers. $d_{\mathrm{model}}=G_{\mathrm{kv}} \cdot d_{\mathrm{head}}$, where $G_{\mathrm{kv}}$ is the number of key-value heads and $d_{\mathrm{head}}$ is the dimension of the attention head.

The representation $\mathbf{z}_{i}$ is further transformed into layer-specific key-value pairs $[\mathbf{C}^{0}_i,\mathbf{C}^{1}_i,\dots,\mathbf{C}_i^{S_{\mathrm{kv}} \cdot L_{\mathrm{kv}}-1}]$, where $\mathbf{C}_i^{j} \in \mathbb{R}^{d_{\mathrm{model}}}$.
For the $l$-th layer, the normalized key-value pair~\cite{onerec_v2} is obtained by: 
\begin{equation}
\label{equ:kv}
\begin{aligned}
\mathbf{k}_i^{(l)} &= \mathrm{RMSNorm}_{k,l}(\mathbf{C}_i^{S_{\mathrm{kv}} \cdot l}) \\
\mathbf{v}_i^{(l)}&= 
\begin{cases}
\mathbf{k}_i^{(l)}, 
& \text{if } S_{\mathrm{kv}} = 1 \ (\text{shared key-value}) \\[4pt]
\mathrm{RMSNorm}_{v,l}\big( \mathbf{C}_i^{S_{\mathrm{kv}}\cdot l +1} \big), 
& \text{if } S_{\mathrm{kv}} = 2 \ (\text{separated key-value}) .
\end{cases}
\end{aligned}
\end{equation}
Finally, the output results for the short-term sequence and long-term sequence are $(\mathbf{K}_\mathrm{short}, \mathbf{V}_\mathrm{short})$ and $(\mathbf{K}_\mathrm{long}, \mathbf{V}_\mathrm{long})$, respectively.

For \textbf{item-dependent behaviors}(e.g., SIM~\cite{sim}), we aggregate the sequence instead of preserving the original sequential structure. The key motivation is to make the key-value representations in cross-attention independent of the target item, thereby enabling request-level inference acceleration (mentioned in Section~\ref{sec:method-optimization-infer}). Specifically, for search-based sequences, we apply a target attention tokenizer to aggregate the sequence into a compact token:
\begin{equation}
\label{equ:search_ta}
\mathbf{e}_ {\mathrm{search}}= \text{TA}(\mathbf{e}_{\mathrm{target}}, \mathbf{K}_{\mathrm{search}}, \mathbf{V}_{\mathrm{search}}),
\end{equation}
where $(\mathbf{K}_{\mathrm{search}}, \mathbf{V}_{\mathrm{search}})$ are the key-value pairs of the search-based sequence, following the same procedure as the aforementioned sequences (Eq.(\ref{equ:kv})).

\subsubsection{Non-sequential Features}\label{sec:method-tokenization-non-seq}
For non-sequential features, we also categorize them into two groups based on their dependency on the target item: \textbf{item-independent features} (\textit{e.g.}, user ID, user profile features, and contextual request features) and \textbf{item-dependent features} (\textit{e.g.}, item ID, item statistics, and user–item cross features). Within each category, features are further grouped according to their semantic meanings, resulting in $m$ item-independent feature groups and $n$ item-dependent feature groups. 
Notably, item-dependent behaviors (e.g., SIM~\cite{sim}) are also treated as item-independent feature and aggregated into a single feature group (\textit{i.e.}, $\mathbf{e}_{\mathrm{search}}$) following Eq.~(\ref{equ:search_ta}).
For each group, the corresponding feature embeddings are concatenated and then projected into a unified hidden space with $d_{\mathrm{model}}$ dimensions through a SwiGLU layer, denoted as $\mathbf{N_{\mathrm{NS}}}\in \mathbb{R}^{q\times d_{\mathrm{model}}}$, where $q = m+n$. 

The rationale for adopting semantic grouping over global grouping~\cite{hyformer}  is two-fold. First, semantic grouping decomposes the computation into user request-related and item-related parts. The former can be computed once per request and reused across candidates, thereby significantly improving inference efficiency. The detailed implementation will be discussed in Section~\ref{sec:method-optimization-infer}. Second, since subsequent interactions are modeled via cross-attention over sequential features, semantically grouped queries provide multiple heterogeneous views, enabling the model to extract diverse user preferences from behavior sequences (shown in Section~\ref{exp:visualization:st}).

\subsubsection{Task Features}\label{sec:method-tokenization-task}
For each task, task-specific features (\textit{e.g.}, task ID, task-related bias features) are grouped into a single token. Similarly, these embeddings are concatenated and projected into a $d_{\mathrm{model}}$-dimensional hidden space through a SwiGLU layer, obtaining $\mathbf{N_{\mathrm{T}}}\in \mathbb{R}^{t\times d_{\mathrm{model}}}$, where $t$ is the number of tasks.

\subsection{Unified Interaction Block} \label{sec:method-interaction}
The unified interaction block jointly integrates an $M$-layer Feature-space Interaction Module (FIM) for modeling interactions between sequential and non-sequential features and an $N$-layer Task-space Interaction Module (TIM) for capturing interactions between high-order features and tasks, enabling unified co-scaling across feature and task spaces.

\subsubsection{Feature-space Interaction Module}\label{sec:method-interaction-feature}
The Feature-space Interaction Module (FIM) is designed to capture interaction patterns within the feature space. It comprises two complementary components: a sequential-oriented interaction component and a non-sequential-oriented interaction component.

\begin{figure}[t]
  \centering
  \includegraphics[width=0.9\linewidth]{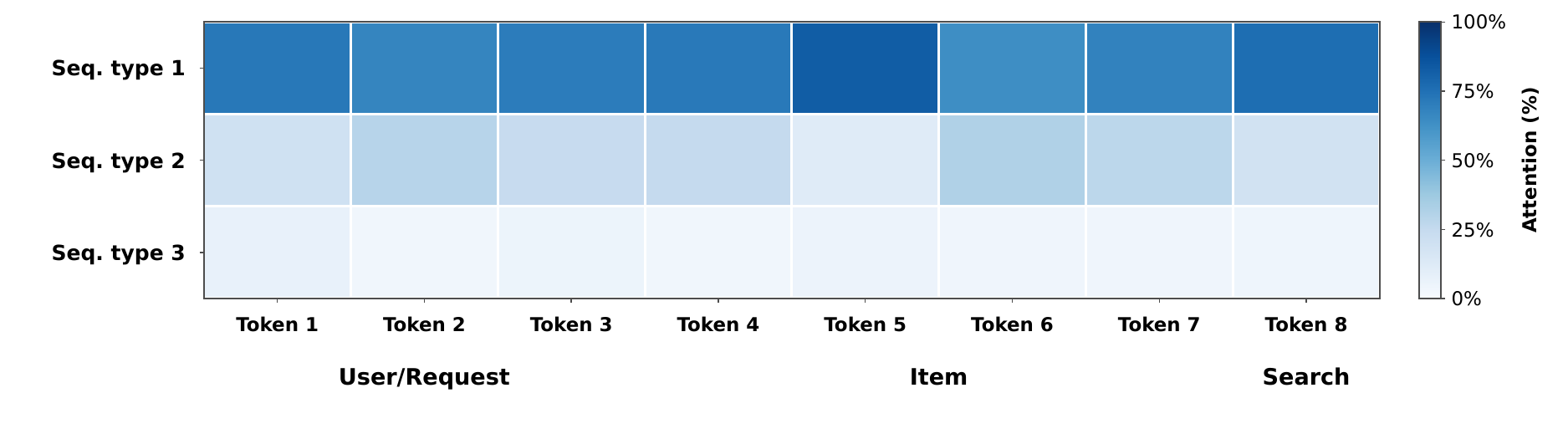}
  \vspace{-6pt}
  \caption{Attention distributions over heterogeneous sequences under shared cross-attention.}
  \label{fig:seq_token_att}
  \vspace{-10pt}
\end{figure}

\textbf{Sequential-oriented Interaction}.  
To effectively extract informative signals from user behavior sequences while preserving both short-term and long-term user interests, we propose a multi-sequence cross-attention mechanism rather than applying a shared cross-attention layer over concatenated heterogeneous sequences.
The key motivation is to avoid attention collapse toward a specific sequence~\cite{gems} (shown in Figure~\ref{fig:seq_token_att}), allowing the model to preserve diverse multi-view interests from heterogeneous behavior sequences.

Taking the short-term sequence as an example, at the $l$-th layer, the previous output tokens are used as queries to perform cross-attention with a residual connection:
\begin{equation}
\mathbf{H}_{\mathrm{short}}^{\mathrm{feat,}(l)}
=
\mathrm{CA}\left(
\mathrm{RMSNorm}(\mathbf{Q}^{\mathrm{feat,}(l-1)}_{\mathrm{cross}}),
\mathbf{K}_{\mathrm{short}}^{(l)},
\mathbf{V}_{\mathrm{short}}^{(l)}
\right) + \mathbf{Q}^{\mathrm{feat,}(l-1)}_{\mathrm{cross}},
\end{equation}
where $\mathbf{Q}^{\mathrm{feat,}(l-1)}_{\mathrm{cross}} \in \mathbb{R}^{q\times d_{\mathrm{model}}}$ denotes the query tokens and $(\mathbf{K}_{\mathrm{short}}^{(l)}, \mathbf{V}_{\mathrm{short}}^{(l)})$ are the $l$-th layer representations derived from the short-term behavior sequence. 
Inspired by the lazy decoder design in OneRec-v2~\cite{onerec_v2}, our key-value (KV) mapping in cross layers also follows the lazy design principle. By default, we set $S_{\mathrm{kv}}=1$ and $L_{\mathrm{kv}}=1$, allowing all layers to share the same KV representations, which significantly reduces memory usage and improves computational efficiency.
Specifically, for the first layer, the $q$ semantically grouped tokens from the tokenization block are used as queries to interact with multiple behavior sequences, \textit{i.e.}, $\mathbf{Q}^{\mathrm{feat,}(0)}_{\mathrm{cross}}=\mathbf{N_{\mathrm{NS}}} \in \mathbb{R}^{q\times d_{\mathrm{model}}}$, enabling the extraction of diverse and complementary information from heterogeneous behavior sources. 
Notably,  to stabilize training, we also adopt a Pre-Norm RMSNorm~\cite{tokenmixer} operation.

The resulting representations are then fed into Sequential-oriented SwiGLU Feed-Forward Networks (S-FFNs) with a residual connection to further enhance sequence-aware patterns:
\begin{equation}
\widetilde{\mathbf{H}}_{\mathrm{short}}^{\mathrm{feat,}(l)}
=
\mathrm{FFN}^{\mathrm{feat,}(l)}_{\mathrm{short}}\left(
\mathbf{H}_{\mathrm{short}}^{\mathrm{feat,}(l)}
\right) + \mathbf{H}_{\mathrm{short}}^{\mathrm{feat,}(l)}.
\end{equation}
Similarly, the long-term behavior sequence is processed in the same manner, producing $\widetilde{\mathbf{H}}_{\mathrm{long}}^{\mathrm{feat,}(l)}$.
Notably, we adopt S-FFNs to model heterogeneous behavior sequences separately. The key rationale is that the Sequential-oriented Interaction component is designed to capture diverse user behavior signals, such as short-term, long-term, and cross-domain sequences. Introducing sequence-specific modeling networks allows \model to better capture heterogeneous behavior patterns within dedicated latent spaces, leading to more fine-grained user interest modeling.

Finally, to integrate the interaction outputs extracted from heterogeneous behavior sequences, we propose two adaptive fusion strategies. 
1) \textbf{Global adaptive fusion} learns a global coefficient $\alpha$ to balance the relative contributions of different behavior sequences: $\mathbf{H}^{\mathrm{feat,}(l)}_{\mathrm{cross}}
=
\alpha \widetilde{\mathbf{H}}_{\mathrm{short}}^{\mathrm{feat,}(l)}
+
(1-\alpha)\widetilde{\mathbf{H}}_{\mathrm{long}}^{\mathrm{feat,}(l)}$, where $\alpha \in [0,1]$ is a learnable global fusion coefficient.
2) Given the substantial differences in short- and long-term behavioral patterns across users, particularly between highly-active and cold-start users, we further introduce \textbf{personalized adaptive fusion}, which dynamically predicts a user-specific fusion coefficient $\alpha_i$ based on user-related features (\textit{e.g.}, activity level and interaction behaviors): $\mathbf{H}^{\mathrm{feat,}(l)}_{\mathrm{cross}}
=
\alpha_i \widetilde{\mathbf{H}}_{\mathrm{short}}^{\mathrm{feat,}(l)}
+
(1-\alpha_i)\widetilde{\mathbf{H}}_{\mathrm{long}}^{\mathrm{feat,}(l)}$. 
%In this way, the model can adaptively balance heterogeneous sequential signals and construct a more comprehensive user representation.

\textbf{Non-Sequential-oriented Interaction}.  
Since the representations $\mathbf{H}^{\mathrm{feat,}(l)}_{\mathrm{cross}}$ extracted by the sequence-oriented interaction are primarily dominated by sequential information, \model further performs \textbf{Interaction Enhancement} by incorporating non-sequential information before sufficient high-order feature interaction modeling. Specifically, $\mathbf{H}^{\mathrm{feat,}(l)}_{\mathrm{cross}}$ is concatenated with the queries from the previous layer $\mathbf{Q}^{\mathrm{feat,}(l-1)}_{\mathrm{self}}$ to enhance feature interaction modeling, \textit{i.e.,} $\mathbf{X}^{\mathrm{feat,}(l)} = [\mathbf{H}^{\mathrm{feat,}(l)}_{\mathrm{cross}} ||\mathbf{Q}^{\mathrm{feat,}(l-1)}_{\mathrm{self}}]$. Specifically, $\mathbf{Q}^{\mathrm{feat,}(0)}_{\mathrm{self}} = \mathbf{Q}^{\mathrm{feat,}(0)}_{\mathrm{cross}}=\mathbf{N_{\mathrm{NS}}}$.
%The representations $\mathbf{H}^{\mathrm{feat,}(l)}_{\mathrm{cross}}$ obtained from the sequence-oriented interaction are concatenated with the queries from the previous layer $\mathbf{Q}^{\mathrm{feat,}(l-1)}_{\mathrm{self}}$, i.e., $\mathbf{X}^{\mathrm{feat,}(l)} = [\mathbf{H}^{\mathrm{feat,}(l)}_{\mathrm{cross}} ||\mathbf{Q}^{\mathrm{feat,}(l-1)}_{\mathrm{self}}]$. Specifically, $\mathbf{Q}^{\mathrm{feat,}(0)}_{\mathrm{self}} = \mathbf{Q}^{\mathrm{feat,}(0)}_{\mathrm{cross}}=\mathbf{N_{\mathrm{NS}}}$.
Subsequently, the concatenated tensor $\mathbf{X}^{\mathrm{feat,}(l)}$ is fed into the non-sequential-oriented interaction component with self-attention and feed-forward layers for interaction modeling.

Specifically,  a self-attention layer with residual connection is applied:
\begin{equation}
\mathbf{H}^{\mathrm{feat,}(l)}_{\mathrm{self}} = 
\mathrm{SA}\left( \mathrm{RMSNorm}(\mathbf{X}^{\mathrm{feat,}(l)}) \right) + \mathbf{X}^{\mathrm{feat,}(l)},
\end{equation}
where $\mathbf{H}^{\mathrm{feat,}(l)}_{\mathrm{self}} \in \mathbb{R}^{2q\times d_{\mathrm{model}}}$. Then, a group of Non-Sequential-oriented SwiGLU Feed-Forward Networks (NS-FFNs) are applied for effective modeling of the feature heterogeneity. Specifically, for each slice $\mathbf{h}^{\mathrm{feat,}(l)}_i, i \in \{0, \dots, 2q-1\}$, a feature-specific FFN is employed to enhance the interactions:
\begin{equation}
\mathbf{f}^{\mathrm{feat,}(l)}_i = 
\mathrm{FFN}^{\mathrm{feat,}(l)}_{i}\left( \mathbf{h}^{\mathrm{feat,}(l)}_i
\right) + \mathbf{h}^{\mathrm{feat,}(l)}_i,
\end{equation}
thus obtaining the output of the $l$-th layer FIM, denoted as $\mathbf{F}^{\mathrm{feat,}(l)} \in \mathbb{R}^{2q\times d_{\mathrm{model}}}$. 
%Subsequently, the tensor $\mathbf{F}^{\mathrm{feat,}(l)}$ is split into two parts, i.e., $\mathbf{Q}^{\mathrm{feat,}(l)}_{\mathrm{cross}} = \mathbf{F}^{\mathrm{feat,}(l)}[:p_\mathrm{cross}]$ and $\mathbf{Q}^{\mathrm{feat,}(l)}_{\mathrm{self}} = \mathbf{F}^{\mathrm{feat,}(l)}[p_\mathrm{cross}+1:]$, which are fed into the cross-attention and self-attention layers of the next layer, respectively.
Subsequently, the tensor $\mathbf{F}^{\mathrm{feat},(l)}$ is partitioned along the feature dimension into two components according to the layer-wise split ratio $\beta^{(l)}$, \textit{i.e.},  $\mathbf{F}^{\mathrm{feat},(l)}=[\mathbf{Q}^{\mathrm{feat},(l+1)}_{\mathrm{cross}}:\mathbf{Q}^{\mathrm{feat},(l+1)}_{\mathrm{self}}]$, where $\mathbf{Q}^{\mathrm{feat},(l+1)}_{\mathrm{cross}} \in \mathbb{R}^{2q\beta^{(l)}\times d_{\mathrm{model}}}$ and $\mathbf{Q}^{\mathrm{feat},(l+1)}_{\mathrm{self}} \in \mathbb{R}^{2q(1-\beta^{(l)})\times d_{\mathrm{model}}}$. The two components are then fed into the next cross-attention and self-attention layers, respectively.
By default, the split ratio $\beta^{(l)}$ is set to $0.5$. In addition, a pyramid-style design can be adopted, where $\beta^{(l)}$ is progressively reduced across layers, resulting in progressively lightweight cross-attention computation and improved efficiency.

\subsubsection{Task-space Interaction Module}\label{sec:method-interaction-task}
After the Feature-space Interaction Module (FIM) performs sufficient interaction modeling in the feature space, \model further introduces the Task-space Interaction Module (TIM) to model the feature-task relationship. Similar to FIM, TIM also comprises two components: a feature-oriented interaction component for intra-task feature interaction and a task-oriented interaction component for inter-task modeling.

\textbf{Feature-oriented Interaction}.  
To perform task-aware feature interaction, \model incorporates a cross-attention layer to perform task-specific interactions over the final feature-space representations $\mathbf{F}^{\mathrm{feat,}(M)}$ extracted by the FIM. Specifically, at the $l$-th layer, the previous outputs are used as queries to perform cross-attention:  
\begin{equation}
\mathbf{H}_{\mathrm{cross}}^{\mathrm{task},(l)}
=
\mathrm{CA}\left(
\mathrm{RMSNorm}(\mathbf{Q}^{\mathrm{task},(l-1)}_{\mathrm{cross}}),
\mathbf{K}_{\mathrm{feat}}^{(l)},
\mathbf{V}_{\mathrm{feat}}^{(l)}
\right) + \mathbf{Q}^{\mathrm{task},(l-1)}_{\mathrm{cross}},
\end{equation}
where  $\mathbf{Q}^{\mathrm{task,}(l-1)}_{\mathrm{cross}} \in \mathbb{R}^{t\times d_{\mathrm{model}}}$ denotes the query tokens and for the first layer, the task feature tokens from the tokenization block are used as queries, \textit{i.e.}, $\mathbf{Q}^{\mathrm{task,}(0)}_{\mathrm{cross}}=\mathbf{N_{\mathrm{T}}} \in \mathbb{R}^{t\times d_{\mathrm{model}}}$. 
Besides, the $(\mathbf{K}_{\mathrm{feat}}^{(1)},
\mathbf{V}_{\mathrm{feat}}^{(1)})$ of the first layer is $\mathbf{K}_{\mathrm{feat}}^{(1)} = \mathbf{V}_{\mathrm{feat}}^{(1)} = \mathbf{F}^{\mathrm{feat,}(M)}$. Similar to FIM, the cross-attention layers in TIM also follow the lazy design by default, sharing key-value representations across all layers to improve efficiency.
This design enables task-aware weighted aggregation over the high-order interaction information $\mathbf{F}^{\mathrm{feat,}(M)}$, providing a multi-task representation analogous to the MMoE framework~\cite{mmoe}.

\textbf{Task-oriented Interaction}.  
Subsequently, to capture inter-task interactions, \model incorporates a self-attention layer followed by a group of Task-oriented SwiGLU Feed-Forward Networks (T-FFNs), further enhancing task-specific modeling capabilities.
\begin{equation}
\begin{aligned}
\mathbf{H}^{\mathrm{task,}(l)}_{\mathrm{self}} &= 
\mathrm{SA}\left( \mathrm{RMSNorm}(\mathbf{H}_\mathrm{cross}^{\mathrm{task,}(l)}) \right) + \mathbf{H}_\mathrm{cross}^{\mathrm{task,}(l)},\\
\mathbf{f}^{\mathrm{task,}(l)}_i &= 
\mathrm{FFN}^{\mathrm{task,}(l)}_{i}\left( \mathbf{h}^{\mathrm{task,}(l)}_i
\right) + \mathbf{h}^{\mathrm{task,}(l)}_i,
\end{aligned}
\end{equation}
where tensor $\mathbf{h}^{\mathrm{task,}(l)}_i, i \in \{0, \dots, t-1\}$ is the slice of $\mathbf{H}^{\mathrm{task,}(l)}_{\mathrm{self}}$ and the output of the $l$-th layer TIM is denoted as $\mathbf{F}^{\mathrm{task,}(l)} \in \mathbb{R}^{t\times d_{\mathrm{model}}}$, which is used as the queries for the next layer, \textit{i.e.}, $\mathbf{Q}^{\mathrm{task,}(l+1)}_{\mathrm{cross}} = \mathbf{F}^{\mathrm{task,}(l)}$.

\subsubsection{Output Module}\label{sec:method-output}

Finally, the output results from the $N$-th layer $\mathbf{F}^{\mathrm{task,}(N)}$ are treated as task-specific representations, which are fed into task-specific FFN heads followed by a Sigmoid activation to produce the final prediction:
\begin{equation}
\hat{y}_i = 
\sigma(\mathrm{FFN}_{i}\left( \mathbf{f}^{\mathrm{task,}(N)}_i
\right)).
\end{equation}
Besides, the weighted multi-task loss over $t$ tasks is adopted to optimize the \model:
\begin{equation}
\mathcal{L}
=
\sum_{i=1}^{t}
\lambda_i
\mathcal{L}_i(y_i, \hat{y}_i),
\end{equation}
where $y_i$ denotes the ground-truth label for the $i$-th task, and $\mathcal{L}_i$ is the task-specific loss such as the Binary Cross-Entropy (BCE)~\cite{multi_task}.

\subsection{Optimization \& Deployment}\label{sec:method-optimization}

\subsubsection{Optimization for Train}\label{sec:method-optimization-train}
To improve training efficiency and scalability, we introduce several system-level optimizations.

\textbf{User-level Common Compression.}  
Generally, multiple candidate items within the same request or session share identical user-side behavior sequences. Due to the multi-valued nature and considerable length of behaviors, redundantly storing and repeatedly processing them will introduce substantial memory and computational overhead. 
Therefore, we apply common compression primarily to user-side behavior features. For each user, the behavior features are stored and processed only once, while an index mapping is used to associate the shared representation with the corresponding item samples. This design reduces redundant sequence embedding lookup, memory copy, and aggregation operations.

%In user-level sampled training, multiple item samples from the same user share identical user-side behavior sequences, while candidate-specific features vary. We therefore apply common compression mainly to user-side sequence features. Each unique user's sequence features are stored and processed once, with an indicator mapping them back to the corresponding item samples, thus reducing the repeated sequence embedding lookup, memory copy, and aggregation.

% For a batch containing \(B\) samples from \(U\) unique users, let \(\bar{k}=B/U\) denote the average number of samples per user. Let \(C_{\mathrm{com}}\) be the cost of processing user-side common features, and \(C_{\mathrm{sample}}\) be the cost of the remaining features. %Without common compression, the cost is
% With user-level common compression, the cost of feature processing can be reduced from \(
% B(C_{\mathrm{com}}+C_{\mathrm{sample}})
% \) to \(
% U C_{\mathrm{com}} + B C_{\mathrm{sample}}.
% \)
% Thus, the theoretical speedup is
% \(
% \frac{B(C_{\mathrm{com}}+C_{\mathrm{sample}})}
% {U C_{\mathrm{com}}+B C_{\mathrm{sample}}}
% =\frac{C_{\mathrm{com}}+C_{\mathrm{sample}}}
% {C_{\mathrm{sample}}+C_{\mathrm{com}}/\bar{k}}.
% \) 
%As \(\bar{k}\) increases, the repeated sequence cost is amortized, and the speedup upper bound approaches
% \[
% S_{\max}=1+\frac{C_{\mathrm{seq}}}{C_{\mathrm{sample}}}.
% \]
% This optimization is most effective when user-side sequence features dominate and each user contributes multiple item samples.

\textbf{Variable-length FlashAttention.}
We adopt variable-length FlashAttention to eliminate padding overhead in heterogeneous user behavior sequences. In UniFormer, semantically grouped query tokens attend to behavior sequences with different lengths. Given a batch of sequence lengths \(\{L_i\}_{i=1}^{B}\) and query length \(q\), padded cross-attention incurs \(O(BqL_{\max}d)\) complexity, while the variable-length formulation reduces it to \(O(q\sum_i L_i d)\). %The ideal padding-removal speedup is therefore
% \[
% S_{\mathrm{varlen}} \approx \frac{B q L_{\max}}{q\sum_i L_i}
% = \frac{B L_{\max}}{\sum_i L_i}.
% \]
Together with FlashAttention's IO-aware tiling~\cite{flashattention}, which avoids materializing the full attention matrix, this optimization reduces both redundant computation and memory traffic.

% \textbf{BF16 Training.}
% Besides, we deploy BF16 mixed-precision training to reduce memory bandwidth and improve throughput. Compared with FP32, BF16 halves tensor storage and enables low-precision tensor cores for matrix-intensive operators, while maintaining sufficient numerical precision for training.

%Compared with FP32, BF16 halves tensor storage and enables low-precision tensor cores for matrix-heavy operators
%. Since BF16 keeps the same exponent width as FP32, it provides better numerical robustness than FP16, leading to more accurate probability estimation for ranking tasks.

\subsubsection{Optimization for Infer}\label{sec:method-optimization-infer}
To reduce online serving latency, the optimization strategy for inference is also adapted.

\textbf{User-Item Decoupling.}
During online serving, one user request is typically paired with \(I\) candidate items for ranking (typically 512 or 1024). Benefiting from the carefully designed tokenization scheme, \model decomposes non-sequential features into \(m\) item-independent tokens and \(n\) item-dependent tokens, while aggregating item-dependent behavior sequences to ensure that the key-value representations in cross-attention remain independent of the target item. %The former remains unchanged within the same request, while the latter varies with each candidate item.
Based on this decomposition, item-independent tokens can be computed once per request and reused across all candidate items, avoiding redundant user-side computation. 

Notably, the cross-attention layers and FFNs in FIM can be naturally decomposed under this tokenization scheme. However, the full self-attention layers involve interactions among all tokens, making user-side representations dependent on candidate items. To address this issue, we apply an attention mask to remove the attention from user-side queries to item-side keys, allowing the user-side attention computation to be performed and reused once per request.
% Under our online serving setup, this optimization yields a 48\% \cb{check} improvement in QPS, while incuring only negligible degradation in both offline AUC and online metrics.

%Empirically, we observe that this approximation incurs only a negligible degradation in AUC.

% Based on this decomposition, item-independent tokens can be computed once per request and reused across all candidate items. For each candidate, \model only computes item-dependent tokens and the necessary user-item interaction layers, avoiding redundant user-side computation. 
% This decoupling may slightly reduce the modeling capacity of early user-item interactions, since part of the user-side computation is performed before seeing candidate-specific information. In practice, this effect is limited because candidate-dependent tokens and downstream interaction layers are still computed per item.

% Let the token-level costs of item-independent and item-dependent processing be proportional to \(m\) and \(n\), respectively. A coupled inference pipeline requires
% \(
% K(m+n)
% \)
% computation, whereas the decoupled pipeline reduces it to
% \(
% m+Kn.
% \)
% The ideal speedup is therefore
% \(
% S_{\mathrm{infer}}=\frac{K(m+n)}{m+Kn}.
% \)
% When \(K\) is large, the speedup approaches
% \[
% S_{\max}=1+\frac{m}{n},
% \]
% indicating larger gains when each request contains many candidates and item-independent computation dominates.

% \newpage

\section{EXPERIMENTS}\label{sec:exp}

\begin{table*}[!t]
  \centering
  \caption{
    Offline results on the KuaiRand-1K and industrial
    datasets. Improvements are computed relative to the first baseline. Boldface denotes the best result, and
    underline indicates the best result among the baselines.
    The \#Params report the number of parameters of the dense
    neural network, and \#FLOPs is measured per batch with a batch size of 2048.
  }
  \label{tab:performance_comparison}
  \vspace{2pt}

  \begingroup
  \setlength{\tabcolsep}{2.2pt}
  \renewcommand{\arraystretch}{1.08}

  \resizebox{\textwidth}{!}{%
  \begin{tabular}{l*{4}{cc}cc|*{4}{cc}cc}
    \toprule
    
    \multirow{3}{*}{Model}
    & \multicolumn{10}{c|}{KuaiRand-1K Dataset}
    & \multicolumn{10}{c}{Industrial Dataset} \\
    
    \cmidrule(lr){2-11}
    \cmidrule(lr){12-21}
    
    & \multicolumn{2}{c}{Effective-view}
    & \multicolumn{2}{c}{Long-view}
    & \multicolumn{2}{c}{Like}
    & \multicolumn{2}{c}{Comment}
    & \multirow{2}{*}{\#Params}
    & \multirow{2}{*}{\#FLOPs}
    
    & \multicolumn{2}{c}{Effective-view}
    & \multicolumn{2}{c}{Long-view}
    & \multicolumn{2}{c}{Like}
    & \multicolumn{2}{c}{Follow}
    & \multirow{2}{*}{\#Params}
    & \multirow{2}{*}{\#FLOPs} \\
    
    \cmidrule(lr){2-3}
    \cmidrule(lr){4-5}
    \cmidrule(lr){6-7}
    \cmidrule(lr){8-9}
    \cmidrule(lr){12-13}
    \cmidrule(lr){14-15}
    \cmidrule(lr){16-17}
    \cmidrule(lr){18-19}
    
    & GAUC & Impr.
    & GAUC & Impr.
    & GAUC & Impr.
    & GAUC & Impr.
    & &
    & GAUC & Impr.
    & GAUC & Impr.
    & GAUC & Impr.
    & GAUC & Impr.
    & & \\
    
    \midrule
    
    SIM+DCN
    & 0.6195 & -
    & 0.6611 & -
    & 0.6406 & -
    & 0.5981 & -
    & 5.02M & 0.32T
    & 0.7418 & -
    & 0.7734 & -
    & 0.8486 & -
    & 0.8361 & -
    & 115.8M & 1.40T \\
    
    SIM+HoME
    & 0.6215 & +0.32\%
    & 0.6629 & +0.27\%
    & 0.6400 & -0.09\%
    & 0.6154 & +2.89\%
    & 5.25M & 0.32T
    & 0.7424 & +0.08\%
    & 0.7740 & +0.08\%
    & 0.8494 & +0.09\%
    & 0.8374 & +0.16\%
    & 114.8M & 1.48T \\
    
    SIM+RankMixer
    & 0.6253 & +0.94\%
    & 0.6661 & +0.76\%
    & 0.6420 & +0.22\%
    & 0.6197 & +3.61\%
    & 15.84M & 0.36T
    & 0.7443 & +0.34\%
    & 0.7757 & +0.30\%
    & 0.8507 & +0.25\%
    & 0.8374 & +0.16\%
    & 492.0M & 3.30T \\
    
    OneTrans
    & 0.6245 & +0.81\%
    & 0.6647 & +0.54\%
    & \underline{0.6567} & \underline{+2.51\%}
    & \underline{0.6247} & \underline{+4.45\%}
    & 17.03M & 2.32T
    & 0.7445 & +0.36\%
    & 0.7760 & +0.34\%
    & \underline{0.8516} & \underline{+0.35\%}
    & \underline{0.8401} & \underline{+0.48\%}
    & 517.0M & 17.10T \\
    
    HyFormer
    & 0.6262 & +1.08\%
    & \underline{0.6678} & \underline{+1.01\%}
    & 0.6518 & +1.75\%
    & 0.6221 & +4.01\%
    & 17.69M & 0.59T
    & 0.7447 & +0.39\%
    & 0.7762 & +0.36\%
    & 0.8512 & +0.31\%
    & 0.8381 & +0.24\%
    & 496.5M & 5.74T \\
    
    MixFormer
    & \underline{0.6269} & \underline{+1.19\%}
    & 0.6673 & +0.94\%
    & 0.6483 & +1.20\%
    & 0.6244 & +4.40\%
    & 16.32M & 0.58T
    & \underline{0.7450} & \underline{+0.43\%}
    & \underline{0.7764} & \underline{+0.39\%}
    & 0.8514 & +0.33\%
    & 0.8388 & +0.32\%
    & 489.4M & 5.96T \\
    
    \midrule
    
    \model
    & \textbf{0.6291} & \textbf{+1.55\%}
    & \textbf{0.6686} & \textbf{+1.13\%}
    & \textbf{0.6600} & \textbf{+3.03\%}
    & \textbf{0.6300} & \textbf{+5.33\%}
    & 16.79M & 0.63T
    & \textbf{0.7457} & \textbf{+0.53\%}
    & \textbf{0.7771} & \textbf{+0.48\%}
    & \textbf{0.8531} & \textbf{+0.53\%}
    & \textbf{0.8435} & \textbf{+0.89\%}
    & 516.0M & 5.82T \\
    \midrule
    \model-Large
    & 0.6306 & +1.79\%
    & 0.6692 & +1.23\%
    & 0.6657 & +3.92\%
    & 0.6341 & +6.02\%
    & 34.90M & 1.09T
    & 0.7465 & +0.63\%
    & 0.7779 & +0.58\%
    & 0.8538 & +0.61\%
    & 0.8448 & +1.04\%
    & 995.3M & 7.78T \\
    
    \bottomrule
  \end{tabular}}
  \begin{tablenotes}[flushleft]
\scriptsize
\item The industrial platform generates \textbf{over 50 billion} interaction logs per day, and a \textbf{+0.05\%} relative GAUC improvement is regarded as highly significant at this scale and can bring substantial business impact~\cite{smes}.
\end{tablenotes}
  \endgroup

  \vspace{-4pt}
\end{table*}

\subsection{Experimental Setting}
\subsubsection{Datasets and Evaluation Protocols}
% 评估指标描述，auc，gauc，logloss等
\label{sec:datasets}
We evaluate \model on the KuaiRand-1K benchmark~\cite{smes} and a large-scale industrial short-video recommendation platform, which serves more than \textbf{400 million daily active users}, generating over \textbf{50 billion interaction logs per day}. 
%匿名 We evaluate \model on the KuaiRand-1K benchmark~\cite{smes} and Kuaishou’s industrial short-video recommendation platform, which is the largest recommendation scenario in Kuaishou and serves more than \textbf{400 million daily active users}, generating over \textbf{50 billion interaction logs per day}.
For KuaiRand-1K benchmark, we evaluate on four tasks: \textit{Effective-view, Long-view, Like}, and \textit{Comment}.
For the industrial dataset, we focus on four key prediction tasks: \emph{Effective-view}, \emph{Long-view}, \emph{Like}, and \emph{Follow} to align with the practical deployment requirements.

We adopt the \textbf{GAUC}~\cite{din,smes} to evaluate the performance of our model, which is the most
critical offline metric in our short-video service. 
%We adopt the widely used ranking metrics to evaluate the predictive performance of our model: \textbf{GAUC}~\cite{zhou2018din}
%In our short-video service, GAUC is the most critical offline metric, which it 
% GAUC calculates the AUC independently for each user and aggregates the per-user values in a sample-weighted manner,
% \begin{equation}
% \mathrm{GAUC} \;=\; \sum_{u} w_u \cdot \mathrm{AUC}_u,
% \qquad
% w_u \;=\; \frac{\#\text{samples}_u}{\sum_{i}\#\text{samples}_i},
% \label{eq:gauc}
% \end{equation}
% where $w_u$ denotes user $u$'s sample proportion in the evaluation set.
Notably, given the stability provided by the massive scale of training samples, a relative GAUC improvement of $\mathbf{0.05\%}$ is regarded as highly significant and can bring substantial business impact~\cite{smes}. To evaluate model efficiency, we additionally report the number of parameters (\#Params) and floating-point operations (\#FLOPs), where FLOPs are calculated with a batch size of 2,048.

\subsubsection{Baselines}
\label{sec:baselines}
We compare \model against several strong representatives that cover both pre-scaling and scaling recommendation models:
%the \emph{sequence-modeling and feature-interaction} separated architecture and the \emph{unified sequence-modeling and feature-interaction} architecture:
\textbf{SIM+DCN}, a classical pre-scaling architecture that combines the Search-based Interest Model (SIM)~\cite{sim} for ultra-long behavior modeling with the Deep Cross Network (DCN)~\cite{dcn} for feature interaction. %, serving as a strong industrial sequence-modeling and feature-interaction reference;
\textbf{SIM+HoME} replaces the DCN component with HoME~\cite{home}, a Hierarchical-Mask MoE module for multi-task supervision.
\textbf{SIM+RankMixer} combines SIM with RankMixer~\cite{rankmixer}, a component-centric scaling module that employs a token-mixing network with scalable parameter capacity. \textbf{OneTrans}~\cite{onetrans}, 
\textbf{HyFormer}~\cite{hyformer} and \textbf{MixFormer}~\cite{mixformer} are cross-module co-scaling methods that jointly scale behavior modeling and feature interaction modeling within the feature space.
% that jointly models behavior patterns and feature interactions in a 
% a unified Encoder backbone, representing a strong unified-encoder reference;
% a transformer-based Hybrid Interaction
% architecture that jointly models feature-space and task-space interactions in
% a unified Encoder backbone, representing a strong unified-encoder reference;
% %and \textbf{MixFormer}~\cite{mixformer}, a recent transformer variant that
% mixes sequence and non-sequence tokens within a single self-attention block,
% representing the state-of-the-art unified sequence/feature interaction
% Encoder design.

\subsubsection{Implementation Details}
% 超参数设置，训练环境
% For a fair comparison, all models were trained from scratch on the same GPU cluster. Specifically, for the industrial dataset, both the sparse and dense components were optimized with AdamW~\cite{adam} using a learning rate of 1e-3 with 1e-3 weight decay, and a per-GPU batch size of 2048. 
% For the industrial dataset, user-side behavioral features include short-term browsing sequences, item-dependent long-term behaviors retrieved based on the target item, and compression-based lifelong interest representations. 
% % For our model \model, it adopts a 3-layer architecture by default, with hidden dimensions $d_{\mathrm{model}}$ of 1280. 
% For models at the 500M scale, we use three layers with $d_{\mathrm{model}}=1280$.
For a fair comparison, all models are trained from scratch on the same GPU cluster. The dense components are optimized using AdamW~\cite{adam} with a weight decay of 1e-3 and a per-GPU batch size of 2048. The learning rate is 1e-3 on the industrial dataset and 1e-4 on KuaiRand-1K.
For KuaiRand-1K, we construct three behavior sequences from users’ histories, corresponding to effective-view, long-view, and like interaction sequences. 
Besides, we employ a two-layer module with \(d_{\mathrm{model}}=180\) and \(d_{\mathrm{head}}=45\).
On the industrial dataset, user-side behavioral features include short-term browsing sequences, item-dependent long-term behaviors retrieved based on the target item, and compression-based lifelong interest representations.
We adopt a three-layer module with \(d_{\mathrm{model}}=1280\) and \(d_{\mathrm{head}}=128\). 
Besides, for both datasets, we adjust the FFN intermediate dimensions to control the overall dense-parameter budget with 17M on KuaiRand-1K and 500M on the industrial dataset.

\subsection{Performance Comparison}
\label{exp:comparision}
We conduct experiments on two datasets against mainstream industrial ranking models, with the results reported in Table~\ref{tab:performance_comparison}. For a fair comparison among scaling-based models, we align their total parameter budgets, using approximately 17M parameters on the KuaiRand-1K dataset and 500M parameters on the industrial dataset. We have the following observations.

%We conduct experiments against mainstream industrial ranking models, and the results are reported in Table~\ref{tab:performance_comparison}. Across all objectives, \model consistently outperforms all state-of-the-art baselines, and we have the following observations.%while UniFormer-Large further achieves the best overall performance.

\begin{itemize}[leftmargin=*]
    \item \textbf{Component-centric scaling.} RankMixer significantly outperforms both DCN and HoME, with GAUC improvements of +0.34\%  and +0.26\% for the \textit{Effective-view} task in the industrial dataset, respectively, demonstrating the effectiveness of parameter scaling in recommendation models.
%RankMixer significantly outperforms both DCN and HoME, demonstrating the effectiveness of parameter scaling in recommendation models. Specifically, RankMixer achieves GAUC improvements of +A\% \cb{calculate} and +B\%\cb{calculate} over DCN and HoME, respectively. %This indicates that the representation capacity of traditional recommendation models is insufficient for complex industrial ranking scenarios, while expanding an individual modeling space can still bring stable gains.

    \item \textbf{Cross-module co-scaling within feature space.} HyFormer and MixFormer further improve performance through cross-module co-scaling, which jointly scales behavior modeling and feature interaction within a unified framework. Their significant gains over SIM+RankMixer demonstrate the necessity of co-scaling, suggesting that scaling multiple components contributes to larger improvements than component-centric scaling.

    % HyFormer and MixFormer further improve performance by co-scaling feature interaction and sequence modeling. Their gains show that modeling non-sequential feature interactions and user behavior sequences in a unified framework can strengthen deep feature interactions, leading to larger improvements than component-centric scaling with Mixer.

    % \item \textbf{Model-centric co-scaling.} \model achieves the best performance among all methods, obtaining a significant improvement over the strongest baselines MixFormer. Unlike HyFormer and MixFormer, which mainly focus on feature-space scaling, \model further incorporates the task space into the unified modeling process. Besides, \model enables flexible and scalable parameter allocation in a model-centric manner, leading to more balanced parameter scaling across the entire model.
    \item \textbf{Model-centric co-scaling.}
    By incorporating the task space into a unified model-centric co-scaling framework, \model enables joint co-scaling across the feature and task spaces, achieving the best overall performance. Compared with feature-space cross-module scaling methods such as HyFormer, MixFormer, and OneTrans, its improvements are particularly pronounced on interaction-oriented objectives (e.g., Like, Comment, and Follow).
    
    % \model achieves the best overall performance, consistently outperforming strong baselines. While HyFormer, MixFormer, and OneTrans explore cross-module co-scaling within the feature space, \model further incorporates the task space into a unified model-centric scaling framework. Moreover, \model supports flexible and scalable parameter allocation across modules and tasks, enabling more balanced scaling of the entire model. 

    % UniFormer achieves the best performance among all compared methods. Unlike HyFormer and MixFormer, which mainly focus on feature-space scaling, UniFormer further incorporates the task space into the unified modeling process. This allows different tasks to actively select their required feature interaction patterns, enabling combinatorial co-scaling among feature interaction, sequence modeling, and task modeling. Compared with the strongest HyFormer/MixFormer baseline, UniFormer still obtains a +C\% GAUC improvement, suggesting that the gain comes from globally better capacity allocation rather than the enhancement of any single module.

    \item \textbf{Further Scaling.}
    \model-Large further improves over \model by +0.24\% and +0.11\% GAUC over two datasets. This improvement can be attributed to the multi-view FFNs tailored to different modeling components, which enable flexible and scalable parameter allocation as the model capacity increases. 
    The parameter-scaling comparison with SOTA baseline is presented in Section~\ref{exp:scaling}.
    % \item 
    % \item Benefiting from the multi-view FFNs designed for different modeling components, \model enables flexible and scalable parameter allocation as the model capacity increases.
    % \item This verifies the favorable scaling property of the proposed framework. The result also demonstrates that UniFormer provides flexible and scalable parameter allocation, enabling the model to simultaneously expand feature representation, sequence understanding, and task-specific modeling capacity.
\end{itemize}

% \begin{table}[H]
%   \centering
%   \caption{Ablation Study of \model on the industrial dataset.}
%   \label{tab:ablation_combined_transposed}
%   \vspace{2pt}
%   \resizebox{\columnwidth}{!}{%
%   \begin{tabular}{lcccc}
%     \toprule
%     \multirow{2}{*}{Variants} &
%     \multicolumn{4}{c}{Industrial Dataset} \\
%     \cmidrule(lr){2-5}
%     & Effective-view & Long-view & Like & Follow \\
%     \midrule
%     \textbf{\model} & \textbf{0.7457} & \textbf{0.7771} & \textbf{0.8531} & \textbf{0.8435} \\
%     \textit{w/o} Multi-seq CA& 0.7455 & 0.7769 & 0.8521 & 0.8431 \\
%     \textit{w/o} IE& 0.7454 & 0.7769 & 0.8523 & 0.8435 \\
%     \textit{w/o} TIM& 0.7453 & 0.7767 & 0.8530 & 0.8434 \\
%     \textit{w/o} S-FFNs& 0.7454 & 0.7769 & 0.8530 & 0.8433 \\
%     \textit{w/o} NS-FFNs& 0.7449 & 0.7764 & 0.8524 & 0.8423 \\
%     \bottomrule
%   \end{tabular}}
%   \vspace{-4pt}
% \end{table}

\subsection{Ablation Study}
\label{exp:ablation}
We conduct an ablation study on the industrial dataset to verify the effectiveness of the key designs of \model. Specifically, we consider the following variants: %replacing Semantic Tokenization with Global Tokenization (\emph{w/o ST}), 
replacing Multi-sequence Cross-Attention with shared Cross-Attention (\emph{w/o Multi-seq CA}), removing Interaction Enhancement (\emph{w/o IE}), \textit{i.e.}, $\mathbf{X}^{\mathrm{feat,}(l)} = \mathbf{H}^{\mathrm{feat,}(l)}_{\mathrm{cross}}$, removing the Task-space Interaction Module (\emph{w/o TIM}), replacing Sequential-oriented Feed-Forward Networks with a shared FFN (\emph{w/o S-FFNs}), and replacing Non-sequential-oriented Feed-Forward Networks with a shared FFN (\emph{w/o NS-FFNs}).

\begin{figure}[t]
  \centering
  \includegraphics[width=\linewidth]{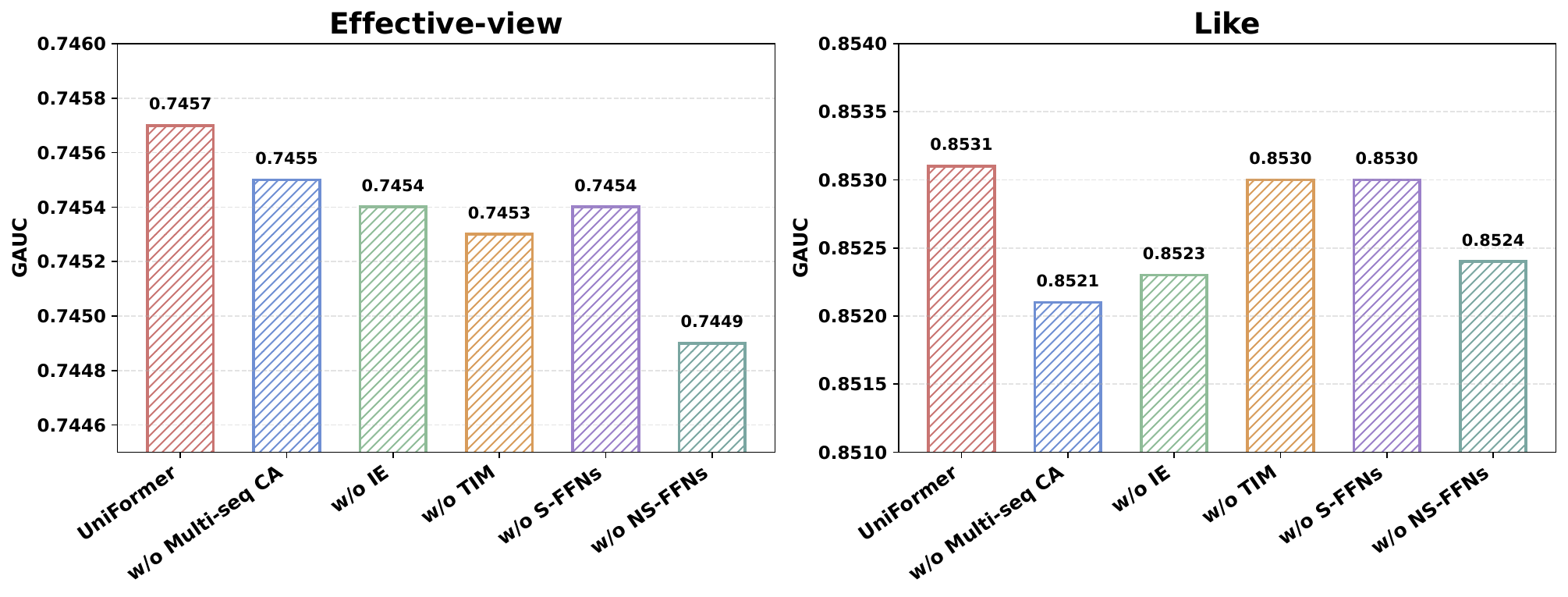}
  \caption{Ablation Study of \model on the industrial dataset over the \textit{Effective-view} and \textit{Like} tasks.}
  \label{fig:ablation}
  \vspace{-10pt}
\end{figure}

\begin{figure}[t]
  \centering
  \includegraphics[width=\linewidth]
  {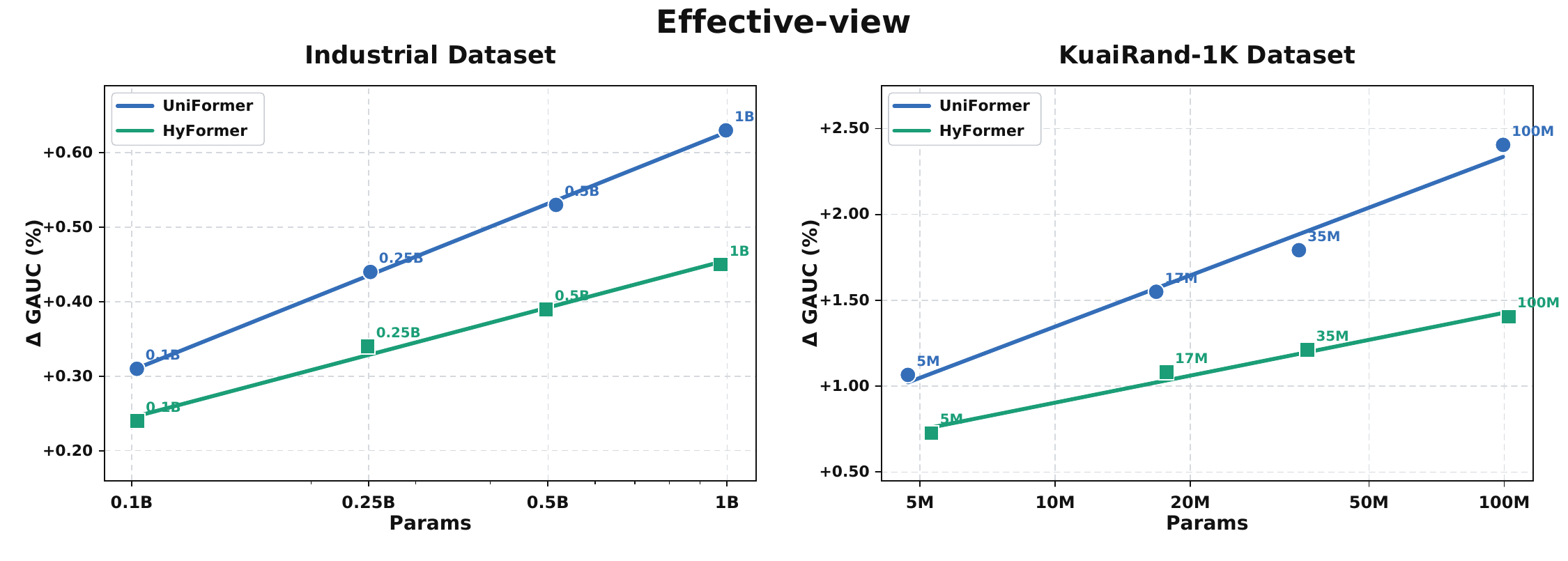}
  % {figure/uniformer_hyformer_scaling_wuauc_1x2_powerlaw.pdf}
  % {figure/decoder_scaling_up_delta_gauc_percent_evtr_like_1x2_powerlaw.pdf}
  \caption{Scaling laws between GAUC gain and model parameters. The x-axis uses a logarithmic scale. }
  \label{fig:scale_up}
  \vspace{-10pt}
\end{figure}

As shown in Figure~\ref{fig:ablation}, \model consistently outperforms all ablated variants across all tasks, demonstrating the effectiveness of the model design.
The results of the first three variants validate the advantages of our unified model design in the Unified Interaction Block. By jointly considering preference completeness, comprehensive interaction modeling, and task relevance, \model achieves more robust and consistent improvements across different tasks.
Notably, removing Multi-sequence Cross-Attention results in a significant performance drop, particularly for interaction-related metrics  (e.g., \textit{Like}), demonstrating the effectiveness of preventing attention collapse.
For the last two variants, replacing S-FFNs or NS-FFNs with a shared FFN leads to the most pronounced performance deterioration. This result demonstrates the effectiveness of our flexible and scalable parameter allocation design, highlighting its crucial role in achieving balanced and effective capacity scaling.

% The \model consistently outperforms every ablation variant in terms of GAUC across all tasks, which verifies the synergistic effect of its core components. \textbf{(1)} Removing ST, Last12Q or TIM leads to significant performance degradation, which indicates that our unified modeling framework benefits substantially from acquiring sequential information from multiple views, complementing the missing raw non-sequential features, and decoupling heterogeneous objectives in the task space. \textbf{(2)} Eliminating Multi-sequence Cross-Attention also incurs a clear accuracy drop, fully validating the necessity of splitting heterogeneous sequences into independent modeling streams to prevent interest collapse. \textbf{(3)} Replacing S-FFNs / NS-FFNs with a shared FFN yields the most pronounced performance deterioration, demonstrating that our flexible parameter-scaling design substantially improves the parameter efficiency of the model.

% 

\subsection{Scaling Analysis}
\label{exp:scaling}
To investigate the capacity potential of \model, we conduct a scaling analysis on the KuaiRand-1K and industrial dataset by increasing the hidden dimensions of multi-view FFNs (including S-FFNs, NS-FFNs, and T-FFNs) and compare with HyFormer, whose results are shown in Figure~\ref{fig:scale_up}. %As observed from the results, the performance of \model consistently improves as the number of parameters increases, exhibiting a clear scaling law trend. 

% As observed from the results, \model exhibits an evident scaling law, where the model performance consistently improves with increasing parameter capacity.
% Moreover, benefiting from the multi-view FFNs tailored to different modeling components, \model enables more balanced and scalable parameter allocation, which helps maintain a stronger performance growth trajectory. 
% Such a model-centric scaling paradigm effectively mitigates the early \textit{Law of Diminishing Returns} often observed in traditional component-centric scaling~\cite{rankmixer}.

% As the model capacity increases, \model consistently achieves greater performance gains on both datasets, exhibiting an approximately linear trend with respect to the logarithmic parameter scale. Moreover, \model maintains a steeper growth trajectory than HyFormer, and their performance gap generally widens as the parameter budget increases. This demonstrates that the additional capacity introduced into \model can be translated into performance improvements more effectively.

% We attribute this favorable scaling behavior to the model-centric co-scaling paradigm of \model. Its multi-view FFNs allocate capacity flexibly across sequential behavior modeling, non-sequential feature interaction, and task-specific modeling, enabling balanced scaling of the entire architecture instead of over-scaling an isolated component. Consequently, \model delays the early performance saturation and mitigates the \textit{Law of Diminishing Returns} commonly observed in conventional component-centric scaling methods~\cite{rankmixer}.[tijian modify]
As the model capacity increases, \model exhibits approximately log-linear performance gains on both datasets and maintains a steeper scaling trajectory than HyFormer, with the performance gap widening at larger parameter budgets. We attribute this behavior to its model-centric co-scaling paradigm, where multi-view FFNs flexibly allocate capacity across sequential, non-sequential, and task-specific modeling. Such balanced whole-model scaling improves the utilization of additional capacity and delays the early saturation commonly observed in component-centric scaling methods~\cite{rankmixer} as the parameter budget grows. 

%%%%%%%%% Scaling数据 evtr     lvtr     ltr      wtr  %%%%%%%%%%
%     Decoder-01B  & 0.7444 & 0.7759 & 0.8515 & 0.8406
%     Decoder-025B & 0.7453 & 0.7767 & 0.8526 & 0.8428
%     Decoder-05B  & 0.7457 & 0.7771 & 0.8531 & 0.8435
%     Decoder-1B   & 0.7465 & 0.7779 & 0.8538 & 0.8448

% \subsection{Efficiency Analysis}
% \label{exp:efficiency}

\subsection{Visualization Analysis}
\label{exp:visualization}
\subsubsection{Semantic Tokenization} 
\label{exp:visualization:st}
\begin{figure}[t]
  \centering
  \includegraphics[width=0.9\linewidth]{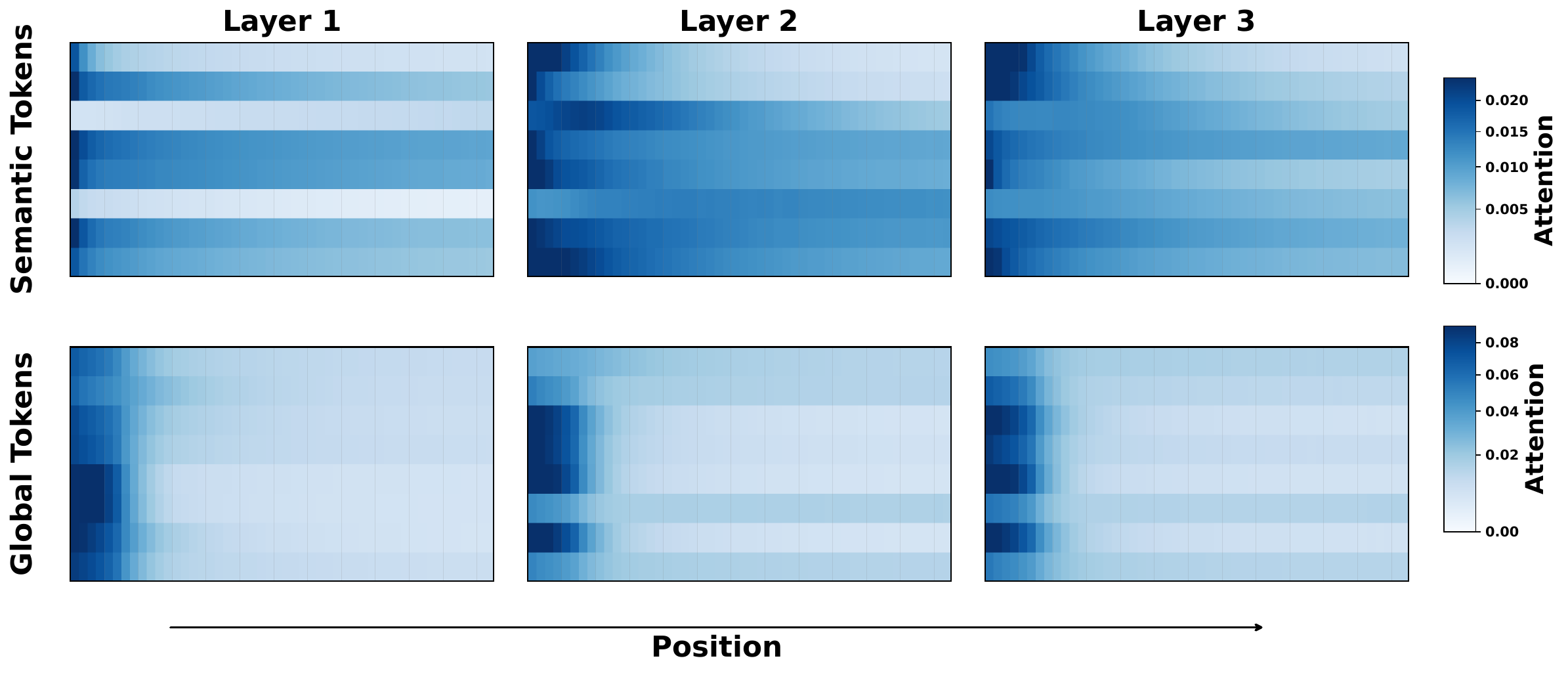}
  \caption{Attention patterns under different tokenization.}
  \label{fig:token_split}
  \vspace{-10pt}
\end{figure}

To illustrate the advantage of semantic tokenization, we compare it with global tokenization~\cite{hyformer,hemix}, which does not explicitly preserve semantic grouping. As shown in Figure~\ref{fig:token_split}, semantic tokens exhibit more diverse and structured attention behaviors across layers for FIM. Across different cross-attention layers, semantic tokens attend to different behavior patterns, enabling \model to capture multi-view user preferences.
In contrast, global tokens exhibit less diverse attention patterns, with attention repeatedly concentrated on similar positions across layers.
Overall, semantic tokenization enables more specialized and layer-adaptive information extraction, thereby mitigating the \textit{Attention Homogenization} caused by global tokenization.

%In contrast, global tokens show much less diversity in their attention patterns. Its attention is consistently concentrated on similar positions across layers, with limited variation from Layer~0 to Layer~2. 
%Overall, this comparison demonstrates that semantic tokenization enables more specialized and layer-adaptive information extraction, while global tokenization tends to produce less differentiated token representations.

% \begin{figure}[t]
%   \centering
%   \includegraphics[width=0.9\linewidth]{figure/attention_layer_pooled_3seq_8tokens_grouped.pdf}
%   \caption{Attention distributions across sequence types.}
%   \label{fig:seq_token}
% \end{figure}

\begin{figure}[t]
  \centering
  \includegraphics[width=0.9\linewidth]{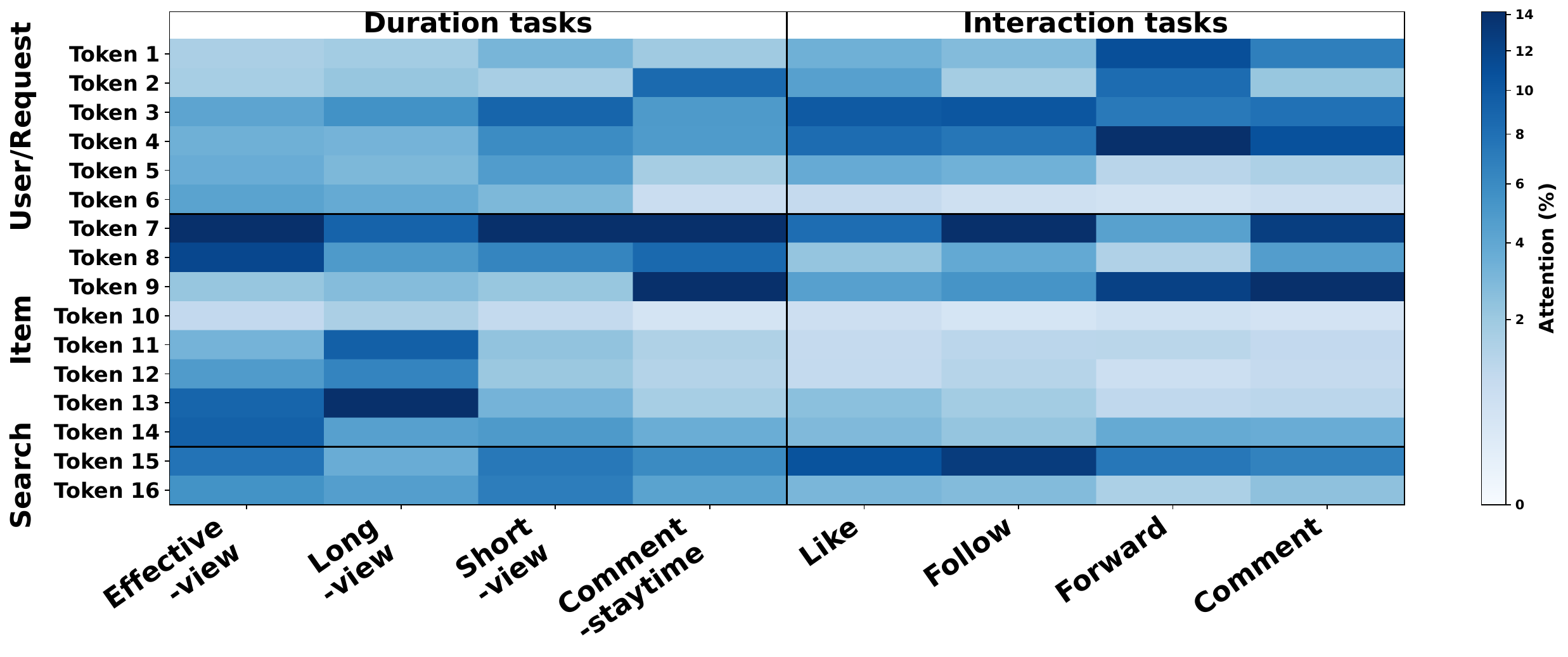}
  \caption{Attention distributions of feature-task relationship.}
  \label{fig:task_token}
  \vspace{-10pt}
\end{figure}

% \subsubsection{Attention Analysis}
% In Figure~\ref{fig:seq_token}, Types I, II, and III denote long-term, medium-term, and recent behavioral interests, respectively. With a single cross-attention (CA) module, query tokens across different types predominantly attend to the long-term sequence, allocating limited attention to the recent sequence. This pattern indicates sequence-level attention collapse. Multi-sequence Cross-Attention alleviates this imbalance by assigning a separate CA branch to each sequence type, with independent softmax normalization. Attention weights therefore sum to one per query within each branch, eliminating competition across sequence types for a shared attention budget. This separation reduces the risk of insufficient attention to slower-converging sequences, supporting more balanced learning across temporal scales while allowing subsequent fusion to determine their relative contributions.

\subsubsection{Feature-Task Relationship} 

To investigate the relationships between features and tasks in TIM,
Figure~\ref{fig:task_token} visualizes the attention distribution of different tasks over feature tokens. 
We observe that different tasks adaptively focus on distinct feature sources according to their specific optimization objectives.
Some features (\textit{e.g.}, item basic features and search-based behavior features) receive consistently high attention weights across different tasks, indicating their important roles in multi-task prediction.
Compared with interaction tasks, item statistical features (Tokens 11-14) receive higher attention weights for duration tasks, whereas certain user features (Tokens 1-4) are more emphasized by interaction tasks.
Moreover, highly related tasks also show strongly aligned attention distributions. For instance, \textit{Comment Staytime} and \textit{Comment Rate} exhibit highly similar attention patterns, suggesting that \model can capture task correlations through feature-task interactions.

% We observe that different tasks selectively focus on different information sources according to their own objectives. 
% For example, some duration-related tasks assign higher attention to item tokens, indicating that content identity, item attributes, and item-level statistics are important for estimating user watch-time objectives. In contrast, several interaction-related tasks show stronger attention to user/request or search-related tokens, suggesting that user-side features and contextual behavioral signals provide complementary information for interaction prediction. 
% For instance, users with stronger interaction preferences, or users who frequently interact with similar items, may require more user-side and search-related signals to accurately model their interactive behaviors.
% Moreover, semantically related tasks tend to exhibit similar token-attention patterns. For example, $cmef$ and $cmtr$ correspond to comment-section dwell time and comment rate, respectively. Although they target different prediction objectives, both are closely associated with comment-related behavior. As shown in the figure, their attention distributions over different token groups are highly similar, which indicates that the model can capture semantic correlations among tasks.

\subsection{Online A/B Testing}
\label{exp:ab}
\subsubsection{Online Results}
\label{exp:ab:result}
% To validate the industrial feasibility of \model, we deployed it on two large-scale short-video recommendation scenarios, for a seven-day online A/B test with $5\%$ of production traffic routed to the treatment bucket. 
% 匿名
To validate the industrial feasibility of \model, we deployed it on the largest short-video channel of our platform Kuaishou and Kuaishou Lite APP, for a seven-day online A/B test with $5\%$ of production traffic routed to the treatment bucket. 
The evaluation metrics include \textit{App Stay Time} and \textit{Watch Time}, which measure the total duration of user engagement, as well as other user interaction metrics, such as \textit{Like} and \textit{Comment}. The online A/B testing results are shown in Table~\ref{tab:online_ab_results}, which demonstrates that \model yields a statistically significant improvement both in user engagement and interaction metrics.
These results demonstrate that our model-centric co-scaling framework delivers consistent improvements across diverse online metrics while effectively mitigating the seesaw effect among competing objectives.

\begin{table}[H]
  \centering
  \caption{Online A/B testing results.}
  \label{tab:online_ab_results}
  \resizebox{0.8\columnwidth}{!}{%
  \begin{tabular}{llcc}
    \toprule
    Category & Metric & Kuaishou Lite & Kuaishou \\
    \midrule
     & App Stay Time & +0.260\% & +0.102\%\\
    Engagement & Watch Time & +1.113\% & +0.729\%\\
    & Video View & +0.252\% & +0.323\%\\
    \midrule
     & Like & +1.089\% & +0.155\%\\
    Interaction & Comment & +1.818\% & +1.488\%\\
     & Collect & +0.930\% & +0.647\%\\
     & Forward & +1.274\% & +0.157\%\\
    \bottomrule
  \end{tabular}}
  \vspace{-4pt}
\end{table}

% \begin{table}[H]
%   \centering
%   \caption{Online A/B testing results. }
%   \label{tab:online_ab_results}
%   \begin{tabular}{llc}
%     \toprule
%     Category & Metric & Relative Impro. \\
%     \midrule
%      & App Stay Time & +0.260\% \\
%     Engagement & Watch Time & +1.113\% \\
%     & Video View & +0.252\% \\
%     \midrule
%      & Like & +1.089\% \\
%      & Follow & +0.214\% \\
%     Interaction & Comment & +1.818\% \\
%      & Collect & +0.930\% \\
%      & Forward & +1.274\% \\
%     \bottomrule
%   \end{tabular}
% \end{table}

\subsubsection{Efficiency Analysis}
\label{exp:ab:efficiency}
Benefiting from our carefully designed tokenization strategy, \model categorizes both sequential and non-sequential features in the feature space into item-independent and item-dependent groups according to their dependency on candidate items. This design enables user-item decoupling during online serving, allowing item-independent network to be computed once per request and reused across multiple candidate items for request-level inference acceleration. In our production serving scenario, each request scores 512 candidate items simultaneously. With user-item decoupling, the inference QPS of \model is improved by \textbf{48\%} compared with the original version, while incurring \textbf{negligible GAUC degradation}.  %These results further demonstrate that \model is well suited for industrial deployment, as it achieves substantial serving acceleration without compromising its strong effectiveness.

\section{conclusion}
In this paper, we advocated a paradigm shift from component-centric scaling to model-centric scaling for industrial recommender systems and systematically discussed the key challenges in building an efficient and unified scaling framework. To this end, we proposed \textbf{\model}, which integrates Feature-space Interaction Modules (FIMs) and Task-space Interaction Modules (TIMs) to jointly scale feature and task modeling. With standardized attention operators and multi-view FFNs, \model enhances modeling capacity while enabling flexible and scalable parameter allocation. Extensive offline experiments and online A/B testing in industrial recommendation scenarios verify the effectiveness and practical value of \model, highlighting model-centric co-scaling as a promising direction for scaling industrial recommender systems.

\balance
\bibliographystyle{ACM-Reference-Format}
\bibliography{main}

% \appendix
% \input{section/appendix}

\end{document}